%% file: ToNDraftJuly7.tex
\documentclass[journal]{IEEEtran}
\usepackage{epsfig, latexsym, amssymb, amsmath,graphics,cite}
\input{./pream}
\input{./math_def}

\begin{document}
\title{Distributed Opportunistic Scheduling With Two-Level Probing}
\author{Chandrashekhar~Thejaswi P.~S.,~\IEEEmembership{Student~Member,~IEEE,}
        Junshan~Zhang,~\IEEEmembership{Senior Member,~IEEE,}
        Man-On~Pun,~\IEEEmembership{Member,~IEEE, }
        H.~Vincent~Poor,~\IEEEmembership{Fellow,~IEEE, }
        Dong~Zheng,~\IEEEmembership{Member,~IEEE}
        \thanks{P.~S.~C.~Thejaswi, J.~Zhang and D.~Zheng are with the Department
of Electrical Engineering, Arizona State University, Tempe, AZ, 85287
e-mail: cpatagup@asu.edu; junshan.zhang@asu.edu; dong.zheng@asu.edu.}
\thanks{M. O. Pun and H. V. Poor are with the Department of Electrical Engineering, Princeton University, Princeton, NJ 08544 e-mail:mopun@ieee.org; poor@princeton.edu.}
\thanks{This research was supported in part
 by the U. S. National Science Foundation under Grants  ANI-0238550,
 CNS-06-25637 and CNS-0721820, and by the Croucher Foundation under a post-doctoral fellowship.}
}
\maketitle

        
\begin{abstract}
Distributed opportunistic scheduling (DOS) is studied for wireless ad-hoc networks in which many links contend for the channel using random access before data transmissions. Simply put, DOS involves a process of joint channel probing and distributed scheduling for ad-hoc (peer-to-peer) communications. Since, in practice, link conditions are estimated with noisy observations, the transmission rate has to be backed off from the estimated rate to avoid transmission outages. Then, a natural question to ask is whether it is worthwhile for the link with successful contention to perform further channel probing to mitigate estimation errors, at the cost of additional probing. Thus motivated, this work investigates DOS with two-level channel probing by optimizing the tradeoff between the throughput gain from more accurate rate estimation and the resulting additional delay. Capitalizing on optimal stopping theory with incomplete information, we show that the optimal scheduling policy is threshold-based and is characterized by either one or two thresholds, depending on network settings. Necessary and sufficient conditions for both cases are rigorously established. In particular, our analysis reveals that performing second-level channel probing is optimal when the first-level estimated channel condition falls in between the two thresholds. Numerical results are provided to illustrate the effectiveness of the proposed DOS with two-level channel probing. We also extend our study to the case with limited feedback, where the feedback from the receiver to its transmitter takes the form of $(0,1,e)$.
\end{abstract}
\section{Introduction}
Channel-aware scheduling has recently emerged as a promising technique to harness the rich diversities inherent in wireless
 networks. In channel-aware scheduling, a joint physical layer (PHY)/medium access control (MAC) optimization is utilized to
  improve network throughput by scheduling links with good channel conditions for data transmissions~\cite{belllab00,qin03,Tse02}. While most existing studies in the literature focus on  centralized scheduling
   (see, e.g.,~\cite{Borst:03,ji-exploiting, Xin:J02, qin03, Tse02}), some initial steps have been taken by the authors to
   develop distributed opportunistic scheduling (DOS) to reap multiuser diversity and time diversity in wireless ad-hoc networks~\cite{zgz:mobihoc}.

 The DOS framework considers an ad-hoc network in which many links contend for the same channel using random access, e.g., carrier-sense multiple-access (CSMA). However, random access protocols provide no guarantee that a successful channel contention is necessarily attained by a link with good channel conditions. From a holistic perspective, a successful link with poor channel conditions should forgo its data transmission and let all links re-contend for the channel. This is because after further channel probing, it is more likely for a link with better channel conditions to take the channel, yielding possibly higher throughput. In this way, multiuser diversity across links and time diversity across time can be exploited in a joint manner. However, each channel probing incurs a cost of contention time. The desired tradeoff between the throughput gain from better channel conditions and the cost for further probing reduces to judiciously choosing an optimal stopping rule for channel probing and the transmission rate for throughput maximization. Using optimal stopping theory (OST), it is shown in~\cite{zgz:mobihoc} that the optimal scheduling scheme turns out to be a pure threshold policy: The successful link proceeds to transmit data only if its supportable rate is higher than the pre-designed threshold; otherwise, it skips the transmission opportunity and lets all other links re-contend. In general, threshold-based scheduling uses local information only and hence it is amenable to easy distributed implementation in practical systems.

 The initial study of DOS~\cite{zgz:mobihoc} hinges upon a key assumption that the channel state information (CSI) is perfectly available at the receiver. In practice, the link rates are estimated with noisy observations. It is shown in~\cite{vakili06} that the signal-to-noise ratio (SNR) estimated by the minimum mean squared error (MMSE) method tends to be larger than the ``actual SNR''. Thus, the transmission rate must be backed off from the estimated rate in order to avoid transmission outages. Our initial steps in \cite{noisydos} show that the optimal scheduling policy under noisy channel estimation still has a threshold structure.

Despite their robust performance under noisy channel estimation, the linear backoff schemes proposed in~\cite{noisydos} back off the rate by a factor that is proportional to the channel estimation errors, which may lead to severe throughput degradation, especially in the low SNR regime, due to a more conservative rate backoff. Recently, wideband communications (e.g., ultra-wideband), has attracted significant attention~\cite{bo_wideband}, owing to its low-power operation and the ability to co-exist with other legacy networks, etc. The great potential of wideband communications gives an impetus to address the problem of throughput degradation, due to estimation errors, in the low-SNR (wideband) regime. More specifically, to circumvent this drawback, a plausible solution is to mitigate the rate estimation errors by performing  further channel probing. In the sequel, we call the initial rate estimation performed {\em during} the channel contention as ``{\em first-level probing}'', whereas the subsequent probing performed {\em after} the successful contention is referred to as ``{\em second-level probing}''. Clearly, the improved rate estimation obtained with second-level probing enables the desired link to make more accurate decisions. However, the advantages of second-level probing come at the price of additional delay. This gives rise to two important questions: 1) Is it worthwhile for the link with successful contention to perform further channel probing to refine the rate estimate, at the cost of additional probing?
2) While there is always a gain in the transmission rate due to the refinement, how much can one bargain with the additional probing overhead?

Specifically, we consider distributed opportunistic scheduling with two-level channel probing. Based on the recent advances in OST, namely OST with two-level incomplete information \cite{Stadje97} and statistical versions of ``prophet inequalities''~\cite{prophet_ineq}, we provide a rigorous characterization of the scheduling strategy that optimizes the tradeoff between the throughput gain achieved by second-level channel probing and the resulting additional delay. It is shown that the optimal scheduling strategy is threshold-based and is characterized by either one or two thresholds, depending on the system parameters. By establishing the corresponding necessary and sufficient conditions for these two cases, we show that the second-level probing can significantly improve the system throughput when the estimated rate via first-level probing falls in between the two thresholds. In such scenarios, the cost of addition delay can be well justified by the throughput enhancement using the second-level channel probing.

 Our intuition is as follows: When the link rate is small, it makes sense to give up the transmission, since the gain due to rate refinement would be marginal due to the poor link condition. On the other hand, when the rate is large enough, it may not be advantageous to perform additional probing as the refinement is meager. Then, it is natural to expect that there exists a ``gray area'' between these two extremes where significant gains are possible by refining the rate estimate with additional probing. We elaborate further on this in Section~\ref{sec:chanest}. Finally, through numerical results, we illustrate the effectiveness of the proposed scheduling scheme.

 Before proceeding further, the main contributions distinguishing this work from other existing works should be emphasized. OST under two levels of incomplete information is addressed with the objective of {\em maximizing the net-return} in~\cite{Stadje97}; in contrast, we study OST with two levels of probing as applied to DOS with the objective of {\em maximizing the rate of return} (i.e., the throughput). Moreover, the optimal strategy proposed in~\cite{Stadje97} presents itself in two mutually exclusive types for decision-making, whereas, our study here reveals that, depending on the parameters, there are more options for DOS, and the corresponding optimal strategy is more sophisticated.
Furthermore, we study distributed opportunistic scheduling for ad-hoc communications under noisy conditions where the rate estimate is available only after a successful channel contention; and this is clearly  different from~\cite{vakili06} which considers centralized scheduling assuming that the rate estimates of all links are available a priori at the base station.
 Despite the fact that both this work and \cite{noisydos} study distributed opportunistic scheduling with imperfect information, this work concentrates on proactively improving throughput by enhancing rate estimation, whereas \cite{noisydos} proposes to passively reduce data rate to avoid transmission outages.
Another related work~\cite{OSTAggregation} uses optimal stopping theory to investigate  the intrinsic
trade-off between energy and delay in distributed data  aggregation and forwarding in sensor networks.

The rest of the paper is organized as follows. In Section~\ref{sec:sysmodel}, we present the system model, and in Section~\ref{sec:first-probing}, we provide background on DOS with only first-level probing in noisy environments. in Section~\ref{sec:chanest}, we present second-level channel probing and characterize the optimal DOS with two-level probing. We also present numerical results to illustrate the gain due to two-level probing. In Section~\ref{limited_feedback}, we extend our study to the case where there is limited feedback from the receiver to its transmitter. Finally, Section~\ref{sec:conclusion} contains our conclusions.

\underline{Notation}: $\left|\cdot\right|$ denotes the amplitude of the enclosed complex-valued quantity. We use $E[\cdot]$ for expectation.

\section{System Model}

\subsection{System model and overview}\label{sec:sysmodel}
\begin{figure}[htp]
\begin{center}
\includegraphics[width=0.2\textwidth]{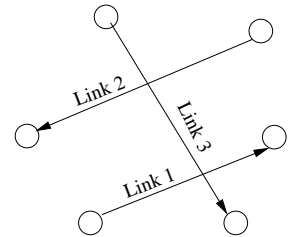}
\end{center}
\caption{Illustration of the ad-hoc network under consideration.} \label{fig:topo}
\end{figure}

Consider a single-hop ad-hoc network in which $L$ links contend for the channel using random access, as illustrated in Fig.~\ref{fig:topo}. A collision model is assumed for random access, where a channel contention of
a link is said to be successful if no other links transmit at the same time.
Let $p_\ell$ be the probability that link $\ell$ contends for the channel, $\ell =1, \ldots, L$. Then the overall successful contention probability, $p_s$, is given by {\small $p_s= \sum_{\ell=1}^L \left(p_\ell \prod_{i \neq \ell} (1-p_i)\right)$}  (cf.~\cite{bertsekas-gallager}).
For ease of exposition, we assume that the contention probabilities, $\{p_\ell\}$, remain fixed (our studies with adaptive contention probability are underway~\cite{weiyan_interference}).
We define the random duration of achieving one successful channel contention as one round of channel probing. Clearly, the number of slots in each probing round, $K$, is a geometric random variable, i.e., $K \sim G(p_s)$. Denoting the slot duration by $\tau$, the corresponding random duration for one probing round thus becomes $K \tau$, with its expected value being $\tau/p_s$.
We assume that a link can estimate its link conditions (hence, the transmission rate) after a successful contention\footnote{The successful link can carry out its rate estimation via a training phase during the request-to-send/clear-to-send (RTS/CTS) handshake, which follows a successful contention. This procedure is fairly standard in the literature, and is not dealt here.}.

Let $s(n)$ denote the successful link in the $n$-th round of channel probing. Due to the nature of wireless channels, the rate in each probing round is random. Following the standard assumption on block fading channels in wireless communications~\cite{sad02}, we assume that the channel remains constant for a duration, $T$ (i.e., $T$ is more or less the channel coherence time). When an estimate of the transmission rate is available, the successful link may decide to transmit over a duration, $T$, if the rate is high enough, or may skip it\footnote{This decision can be broadcast to all users in the one-hop neighborhood (e.g., NCTS).} and allow all links to re-contend, in the hope that another link with a better channel will take the channel later.

\begin{figure}[htp]
    \centering
    \includegraphics[width=0.45\textwidth]{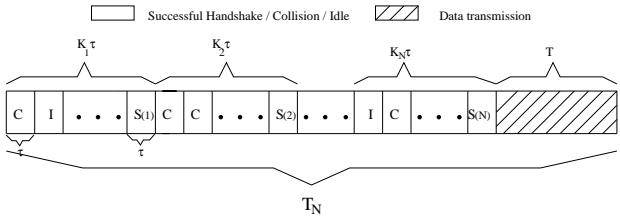}
    \caption{A sample realization of channel probing and data transmission.}
    \label{fig:sample_DOS}
\end{figure}

To get a more concrete sense of joint channel probing and distributed scheduling, we depict, in Fig.~\ref{fig:sample_DOS}, an example with $N$ rounds of channel probing and one single data transmission. Specifically, suppose after the first round of channel probing with a duration of $K_1$ slots, the rate of link $s(1)$ is very small (indicating a poor channel condition); and as a result, $s(1)$ gives up this transmission opportunity and lets all links re-contend. Then, after the second successful contention with a duration of $K_2$ slots, link $s(2)$ also gives up the transmission because its rate is also small. This continues for $N$ rounds until link $s(N)$ transmits because its transmission rate is good. Clearly, there exists a tradeoff between the throughput gain from better channel conditions and the cost for further probing.

In~\cite{zgz:mobihoc}, we show that the process of joint channel probing and distributed scheduling can be treated as a team game in which all links collaborate to {\em  maximize rate of return} (the average throughput).
Specifically, as illustrated in Fig.~\ref{fig:sample_DOS}, after one round of channel probing, a stopping rule $N$ decides whether the successful link carries out data transmission, or simply skips this opportunity to let all links re-contend.
Let $R_n$ be the transmission rate of the successful link in the $n$-th round of channel probing, and $T_n=\sum_{j=1}^nK_j\tau+T$ be the total system time, defined as the sum of the contention time and the transmission duration.
It turns out that the optimal DOS strategy achieving the maximum throughput hinges on the optimal stopping rule $N^*$, which yields the maximal rate of return $\theta^*$. That is,
\begin{equation}
\theta^* \triangleq \sup_{N\in Q} \frac{E[R_NT]}{E[T_N]},
\label{main-problem}
\end{equation}
and
\begin{equation}
N^* \triangleq \argmax\limits_{N\in Q} \frac{E[R_N T]}{E[T_N]},
\end{equation}
where
\begin{equation} \label{eq:q}
Q \triangleq \{N: N\ge 1, E[T_N]<\infty\}.
\end{equation}

It is clear that $R_n$ plays a critical role in distributed opportunistic scheduling. In practice, rate estimates are seldom perfect. It is shown in~\cite{vakili06} that the rate corresponding to the estimated SNR tends to be greater than the actual rate, and subsequently the transmission rate must be backed-off from the estimated rate to avoid outages. Then, a natural question to ask is whether it is worthwhile for the link with successful contention to perform further channel probing to refine the channel estimate, at the cost of additional probing overhead, and how much can one bargain?

Intuitively speaking, when the transmission rate is small, it makes sense to give up the transmission, since the gain due to rate refinement would be marginal due to the poor link conditions. On the other hand, when the rate is large enough, it may not be advantageous to perform additional probing as the improvement is meager. It is natural to expect that there exists a ``gray area'' between these extremes where significant gains are possible by refining the rate estimate with additional probing. In what follows, we seek a clear understanding of the above fundamental issues.

To this end, we present the PHY-layer signal model first. The received signal corresponding to $s(n)$ can be written as\footnote{We note that the results reported here can be  extended to frequency-selective fading channels by replacing scalar fading parameters with vectors.}
\begin{equation} \label{eq:channel}
Y_{s(n)}(n) = \sqrt{\rho}h_{s(n)}(n)X_{s(n)}(n) + \xi_{s(n)}(n),
\end{equation}
where $\rho$ is the {\em normalized} receiver SNR, $h_{s(n)}(n)$ is the channel gain for link $s(n)$, $X_{s(n)}(n)$ is the transmitted signal with {\small $E[\left|X_{s(n)}(n)\right|^2]=1$}, and $\xi_{s(n)}(n)$ is additive white Gaussian noise (AWGN) with unit variance. In this work, we consider a homogeneous network in which all links are subject to independent Rayleigh fading, with identical channel statistics. Moreover, $h_{s(n)}(n)$ and $h_{s(m)}(m)$ are independent for $n\neq m$, and remain constant over the channel coherence time. This is a practically valid assumption because the likelihood of one
link (say link $m$) achieving two consecutive successful channel probing, $p_m^2 \sum_{i\neq m} (1 - p_i)^2$, is fairly small, especially when the number of links in the network is large. Furthermore, even
if the same link successfully obtains two consecutive channel
contentions, the channel conditions corresponding to the two consecutive
successful channel probings are independent since the channel
probing duration in between is designed to be comparable to
the channel coherence time.

Without loss of generality, we focus on the $n$-th probing round and omit the temporal index $n$, whenever possible, for notational simplicity. For presentational simplicity, we use $Y_n$, $X_n$ and $\xi_n$ and $h_n$ to denote $Y_{s(n)}(n), X_{s(n)}(n)$, $\xi_{s(n)}(n)$ and $h_{s(n)}(n)$, respectively, in the sequel.

When perfect CSI is available to the source node as assumed in \cite{zgz:mobihoc}, the instantaneous supportable data rate is given by the Shannon channel capacity:
\begin{equation}
R_n =W \log(1+\rho |h_n|^2),
\end{equation}
where $W$ is the bandwidth, and $\left\{R_n,\;n=1,\ldots,\right\}$ are independent due to the independence assumption on $h_n$.

  To facilitate our analysis, we concentrate our following investigation in the low SNR (wideband) regime, assuming $\rho\rightarrow 0$ and $W=\Theta(\frac{1}{\rho})$. It is well known that a decrease in SNR estimation error can only increase the rate of communication. For cases with wideband signaling (e.g. in the low
SNR regime), where an increase in the SNR results in a linear increase in the throughput, obtaining
more accurate estimates of the SNR can yield substantial benefits.

\subsection{DOS with one-level probing}
\label{sec:first-probing}
In this section, we briefly examine DOS with one-level channel probing in the low SNR regime~(cf.\cite{noisydos}). Let $M$ be the training length. Thus $\tau_t=MT_s,$ where $T_s$ is the symbol duration. We assume that $M$ is large enough to ensure that $\tau_t=\Theta (1)$.
Assume that the rate estimation is performed based on the MMSE principle.
For simplicity, we assume that the probing symbols $X_m=1$, without considering pilot design. Then, it is straightforward to show that, $\hat{h}_n^{(1)}$, the minimum mean square error (MMSE) estimate of $h$ is given by~\cite{bl:kay}
\begin{equation} \label{eq:hatnLS}
\hat{h}_n^{(1)} = \frac{\sqrt{\rho}}{\rho M +1} \sum_{m=1}^{M} Y_m,
\end{equation}

Substituting~(\ref{eq:channel})~into~(\ref{eq:hatnLS}), we can express $h$ in terms of $\hat{h}^{(1)}$ as
\begin{equation} \label{eq:hnM}
h = \hat{h}_n^{(1)} + \tilde{h}_n^{(1)},
\end{equation}
where $\tilde{h}_n^{(1)}$ is the corresponding estimation error.
Note that the orthogonality principle holds, and we can verify that {\small $\tilde{h}_n^{(1)}$} and {\small $\tilde{h}_n^{(1)}$} are uncorrelated,
where

\begin{equation} {\hat h}_n^{(1)}\sim {\cal CN}\left(0,\frac{\rho M}{\rho M+1}\right)\end{equation}
and
\begin{equation} \tilde{h}_n^{(1)}\sim {\cal CN}\left(0,\frac{1}{\rho M+1}\right).\end{equation}

Without perfect CSI, the source node employs the {\em estimated} SNR $\{\rho|\hat{h}_n^{(1)}|^2, n=1,\ldots\}$ as the basis for distributed scheduling, despite the fact that the {\em actual} SNR is given by
\begin{equation} \label{eq:acsnr}
\lambda_n^{(1)} = \frac{\rho|\hat{h}_n^{(1)}|^2}{1+\rho|\tilde{h}_n^{(1)}|^2},
\end{equation}
where we have modeled the channel estimation error as additive Gaussian noise~\cite{vakili06}.

Inspection of (\ref{eq:acsnr}) reveals that $\lambda_n^{(1)}$ is always {\em smaller} than the estimated SNR  $\{\rho|\hat{h}_n^{(1)}|^2\}$, in the presence of channel estimation errors. As a result, an outage occurs if the source node transmits at a data rate specified by $\{\rho|\hat{h}_n^{(1)}|^2\}$. To circumvent this problem, a linear backoff scheme is used to reduce the data rate~\cite{noisydos}. More specifically, the estimated SNR is linearly backed off to $\sigma_M\rho|\hat{h}_n^{(1)}|^2$, where $\sigma_M$ is the backoff factor with $0<\sigma_M<1$.
Under imperfect information, the transmission rate in the low-SNR wideband region simplifies to
\begin{equation}
\label{eq:RM}
{R}_n^{(1)}  \approx \rho W\sigma_M |\hat{h}_n^{(1)} |^2.
\end{equation}
For more details of arriving at the above equation, we refer to Appendix~\ref{rate}.
It can be shown that the optimal DOS policy with noisy channel estimation remains threshold-based with the optimal threshold $\hat \theta$ given as the solution to the following optimality equation (cf.~\cite{noisydos}):
\begin{equation}
\label{optimality_2}
E\left[{R}_n^{(1)}-{\theta }\right]^+=\frac{\theta\tau}{p_s T}.
\end{equation}
\begin{figure}[t]
\begin{center}
\vspace{0.3cm}
\includegraphics[width=0.42\textwidth]{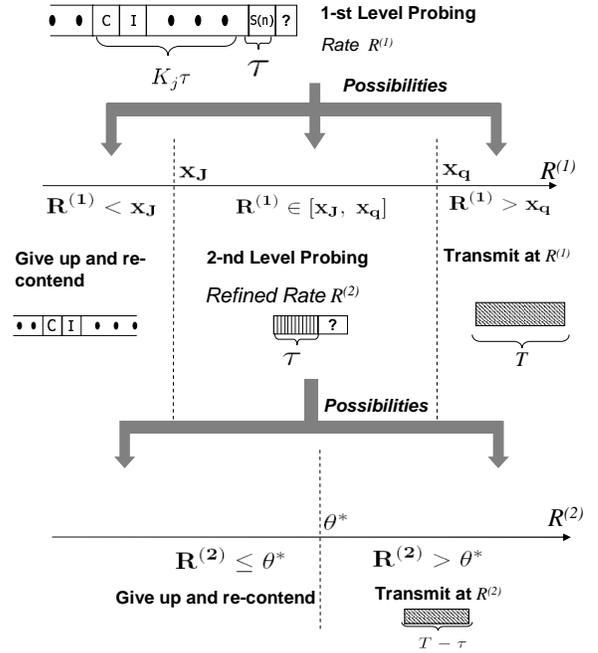}
\end{center}
\caption{A sketch of DOS with two-level probing.}
\label{fig:two_level}
\end{figure}
%

\section{DOS with Two-Level Probing} \label{sec:chanest}
In this section, we characterize the optimal DOS with two-level probing, i.e., the links may choose to refine their rate estimates before making a decision on whether to transmit or not. We illustrate, in~Fig.~\ref{fig:two_level}, the underlying rationale behind DOS with two-level probing. In the following, we detail the procedure with second-level probing, and then cast DOS with two-level probing as a problem of maximal rate of return using optimal stopping theory with incomplete information. We then characterize the corresponding structure and provide a complete description of the optimal strategy.

\subsection{Second-level probing}
\label{subsec:second_level}
 To improve the estimation accuracy, the receiver of the successful link can request its transmitter to send another pilot packet, at the cost of a part of the data transmission duration allotted to it. More specifically, in addition to the pilot symbols sent during the first-level probing, the receiver obtains a refined estimate of $h_n$ by exploiting the newly transmitted pilot symbols of length $\tau_t$ during second-level probing (of duration $\tau$). Then, the link uses the remaining $T-\tau$ for the data transmission. We let ${\hat h}_n^{(2)}$ denote this refined estimate of $h_n$, obtained via two-level probing.
  Following the same procedure described in the previous section, we can show that
\begin{equation} \label{eq:hn2M}
\hat{h}_n^{(2)} = \frac{\sqrt{\rho}}{2\rho M +1} \sum_{i=1}^{2M} Y_i,
\end{equation}
We can show that the estimate ${\hat h}_n^{(2)}$, and the corresponding estimation error $\tilde{h}^{(2)}$ are uncorrelated, where
\begin{equation} {\hat h}_n^{(2)}\sim {\cal CN}\left(0,\frac{\rho 2M}{\rho 2M+1}\right)\end{equation}
and
\begin{equation} \tilde{h}_n^{(2)}\sim {\cal CN}\left(0,\frac{1}{\rho 2M+1}\right).\end{equation}

Finally, the resulting data rate is computed as
\begin{equation}
R_n^{(2)}=\rho W \sigma_{2M} |{\hat h}_n^{(2)}|^2,
\end{equation}
where $\sigma_{2M}$ is the corresponding linear rate back-off factor.

Next, we establish the relationship between the estimates due to first-level and second-level probings. To this end, we apply the principle of linear estimation to represent $\hat{h}_n^{(2)}$ as the linear combination of orthogonal components $h_e$ and $ \hat{h}_n^{(1)}$ as
\begin{equation}
\label{linest}
\hat{h}_n^{(2)} = \hat{h}_n^{(1)} + h_e,
\end{equation}
where $h_e \sim \mathcal{CN}(0,\sigma^2_e)$, with {\small $\sigma^2_{e} = \frac{M\rho}{(M\rho+1)(2M\rho+1)}.$}

By orthogonality, we have
\begin{equation}
E[|\hat{h}_n^{(2)}|^2] = E[|\hat{h}_n^{(1)}|^2] + \sigma^2_{e}.
\end{equation}
For convinience, we introduce a new random variable $z=\sqrt{\sigma_{2M} W\rho}h_e$, where $z\sim {\cal CN}(0,R_e)$, with $R_e=\sigma_{2M} W\rho \sigma_e^2$.
Thus, it follows that the expected rate of the second-level probing conditioned on the rate due to first-level probing, obeys the following relationship:
\[E[R_n^{(2)}|R_n^{(1)}]=c_rR_n^{(1)}+R_e.\]
 We note that $R_e$ can be interpreted as the expected relative rate gain due to the second level probing.

%

\subsection{Scheduling options and rewards}
In what follows, we devise DOS with two levels of probing using optimal stopping theory. Drawing on the ideas from~\cite{ferguson:os}, we show that optimizing the network throughput via DOS can be cast as a {\em maximal rate of return} problem.

Consider the network model in Fig.~\ref{fig:topo}. It takes a total duration of {\small $\sum_{j=1}^n K_j\tau$} to reach the $n$-th round of probing. After the $n$-th round of probing, the successful link has the following three options after computing its rate $R^{(1)}_n$:
\begin{enumerate}
 \item Transmit at rate $R^{(1)}_n$ for time $T$;
 \item Defer transmission and let all nodes re-contend;
 \item Perform second-level probing to obtain the new rate $R^{(2)}_n$, and then decide whether to transmit at $R^{(2)}_n$ for time $T-\tau$ or to defer and re-contend.
\end{enumerate}

Clearly, the basis for distributed opportunistic scheduling with two-level probing is the observation sequence
{\small$\left\{R^{(1)}_n, R^{(2)}_n\right\}_n$}, with the option of skipping {\small$R^{(2)}_n$}. We introduce $\theta$ as the {\em Lagrange multiplier}, which can be interpreted as the cost for using one unit of the system time (that includes the contention, probing and transmission times). Then, the first successful channel probing incurs an average cost of $\theta\tau/p_s$, the second-level probing incurs $\theta\tau$, whereas the data transmission for a duration of $T_d$ entails a cost $\theta T_d$.

Let {\small {$\displaystyle\phi_n:{\cal R}^+ \rightarrow \{0,1,2\} \mbox{ and }\psi_n: {\cal R}^+\rightarrow \{0,1\}$}} be the decision sequences after $R_n^{(1)}=x$ is observed. In particular, $\phi_n(x)=1$ refers to transmitting at the current rate, $\phi_n(x)=0$ means giving up the transmission and re-contend, while $\phi_n(x)=2$ indicates engaging in the second-level probing. Furthermore, when $\phi_n(x)=2$, the final decision hinges on $R_n^{(2)}=y$: if $\psi_n(y)=1$, the link transmits at the refined rate, whereas if $\psi_n(y)=0$, the link gives up the transmission and lets all nodes re-contend.

 Next, let $N$ denote the stopping rule
\[N=\inf\{n\ge 1|  \phi_n=1, \mbox{ or }  \phi_n =2 \mbox{ and } \psi_n=1  \}.\]
Then, the expected net reward ({\em expected return}) is given by
{\small\begin{eqnarray}
r&=& E\left[ R_NT_{d,N}-\theta T_N \right],
\end{eqnarray}}
where $R_n$ is the transmission rate after the $n$-th probing round and is given by
\[R_n=I(\phi_n=1)\cdot R_n^{(1)}+I(\phi_n=2)I(\psi_n=1)\cdot R_n^{(2)},\]
where $I(\cdot)$ is the indicator function, and $T_n$ is the total time given by
\[T_n=\sum_{j=1}^n K_j\tau + T,\]
 which defined as the sum of total contention time (and time due to second-level probing, when performed) and the data transmission duration, {\small $T_{d,n}= T-I(\phi_n=2)I(\psi_n=1)\tau$}.
The corresponding {\em rate of return} is {\small$E[R_NT_{d,N}]/E[T_N]$.}
The maximal expected return is given by
\begin{eqnarray}
r_0&=& \sup_{N\in{Q}}E\left[ R_NT_{d,N}-\theta T_N \right].
\end{eqnarray}
Note that the expected return, $r$, depends on the decision functions $\phi$, $\psi$, and the cost $\theta$.
The principal objective is to maximize the rate of return (i.e., the throughput) of the DOS with two-level probing, defined as
\[\theta^*=\sup_{N\in {Q}}\frac{E\left[ R_NT_{d,N}\right]}{E\left[T_N \right]}.\]

Summarizing, we are interested in seeking a stopping rule $N\in Q$ that obtains $\theta^*$.
The following lemma relates the optimal throughput $\theta^*$ to the expected optimal return $r_0$, and guarantees the existence of such an optimal stopping rule.
\newtheorem{coro}{\em Corollary}
\newtheorem{theorem}{\bf Theorem}
\newtheorem{lemma}{\bf Lemma}
\begin{lemma}
\label{exist}
{ \em For DOS with two-level probing, the optimal stopping rule $N^*$ exists. Furthermore,
$\theta^*$ is attained at $N^*$, and $\theta^*$ satisfies
\[r_0=\sup_{N\in {Q}}E\left[ R_NT_{d,N}-\theta^* T_N \right]=0,\]
 }
\end{lemma}
\proof See Appendix~\ref{app:exist}.

Next, we derive the optimality equation for DOS with two-level probing. Without loss of generality, we assume the transmission duration to be unity, i.e. $T=1$.

We begin by considering the option of second-level probing and introducing its associated reward function. Suppose after observing $R^{(1)}_n=x$, the link performs a second-level probing to obtain $R_n^{(2)}$, and then uses an optimal strategy thereafter. Then, depending on $R^{(2)}_n=y$, it may choose to transmit at rate $y$, for a duration of $1-\tau$, if the associated reward is greater than $r$ (the expected net reward); otherwise it would defer and re-contend.
The reward associated with the data transmission is $(1-\tau)y-(1-\tau)\theta$ in this case. In a nutshell, the expected net reward corresponding to the second-level probing is then given by

{\small\begin{eqnarray}
\nonumber
J_{\theta}(x,r)&\triangleq& rG(\frac{r}{1-\tau}+\theta|x)\\ \label{eq:hxr}
&&+(1-\tau)\int_{\frac{r}{1-\tau}+\theta}^\infty\hspace{-2mm}
(y-\theta)G(dy|x) - \theta\tau,
\end{eqnarray}
}
{\noindent}where $G(y|x)$ is the conditional cumulative distribution function (cdf) of $R_n^{(2)}$, given $R_n^{(1)}=x$. Note that $G(y|x)$ is non-central $\chi^2$ with two degrees of freedom. Furthermore, both $R_n^{(1)}$ and $R_n^{(2)}$ are exponentially distributed. We use $F$ and $F_1$ respectively, to denote the cdfs of $R_n^{(1)}$ and $R_n^{(2)}$.
Finally, it can be shown that $\displaystyle\lim_{x\rightarrow \infty}G(y|x)=0$ and $E\left[y|x\right]=c_rx+R_e$.

In a nutshell, upon observing $R_n^{(1)}=x$ after the $n$-th probing round, the link $s(n)$ can obtain one of the following three rewards:

\begin{enumerate}
\item  $x-\theta$: the reward by transmitting at a rate $x$;
\item $r_0$: the reward obtained by forgoing the current opportunity and re-contending;
\item $J_{\theta}(x,r_0)$: the reward by resorting to refining the rate via second-level probing.
\end{enumerate}
Note that, in computing the rewards above, we have omitted the cost for obtaining the first successful channel probing, i.e. $\theta\tau/p_s$, since it is common to all three returns. The optimal strategy for the link is to choose the option that yields the maximum of the above rewards.
It follows that the optimal DOS strategy with two-level probing satisfies the following optimality equation~ \cite{Stadje97}:
\begin{equation}
\label{eq:optEmax1}
E\left[\max\left\{R^{(1)}-\theta, r_0, J_{\theta}(R^{(1)},r_0)\right\}\right]-\frac{\theta\tau}{p_s} = r_0,
\end{equation}
where $R^{(1)}$ has same distribution as $R_n^{(1)}$.

From Lemma~\ref{exist}, when the throughput, as a function of $\theta$, reaches its maximum, we have that $r_0=0$ at $\theta=\theta^*$ \cite{ferguson:os}. Thus, (\ref{eq:optEmax1}) can be rewritten as
\begin{equation}\label{eq:optEmax2}
E\left[\max\left\{R^{(1)}-\theta^*,J_{\theta^*}(R^{(1)},0)\right\}\right]^+=\frac{\theta^*\tau}{p_s}.
\end{equation}
Inspection of (\ref{eq:optEmax2}) indicates that the second-level probing is optimal when $J_{\theta^*}(x,0)>0$ and $J_{\theta^*}(x,0)>x-\theta^*$ for some $x$.

It is worth noting that the following fact holds:
\begin{equation}
\label{fact}
\theta^*>\theta_L\stackrel{\Delta}{=} \frac{E[R^{(1)}]}{\frac{\tau}{p_s}+1}.
\end{equation}
Note that $\theta_L$ corresponds to the throughput of {\em PHY-Oblivious scheduling},
which is a single-level probing scheme with zero threshold.
This can be achieved by the degenerate stopping rule, which stops at the very first time.

\subsection{Structure of optimal scheduling strategy}
We next proceed to study the structure of the optimal scheduling strategy. Essentially, the optimal strategy takes a threshold form. Depending on the specific network setting, the optimal strategy may admit one of the two intuitively reasonable types, namely Strategy A and Strategy B.
Generally speaking, under Strategy A, it is always optimal to demand additional information when the estimated rate lies between two thresholds. This is the case when the gain due to second-level probing is comparable with the additional overhead. In contrast, under Strategy B, there is never a need to appeal to second-level probing. This case occurs for example, when the improvement due to the refinement is dominated by the probing overhead.
An extreme example of this case is when perfect CSI is available to the transmitter.

Before we state the main result on the optimal strategy, we define $q(x)\stackrel{\Delta}{=}J_{\theta^*}(x,0)-x+\theta^*$. Intuitively speaking, $q(x)$ represents the expected gain achieved by second-level probing compared to directly transmitting at the current rate. Thus, if $q(x)>0$, performing second-level probing is a better option than directly proceeding to data transmission. We need the following lemmas before characterizing the structure of the optimal scheduling strategy.

\renewcommand{\theenumi}{\roman{enumi}}

\begin{lemma}\label{h_fn}
$J_{\theta^*}(x,0)$ and $q(x)$ are characterized by the following properties:
\begin{enumerate}
\item $J_{\theta^*}(x,0)$ is monotonically increasing in $x$ with $\displaystyle\lim_{x\rightarrow\infty} J_{\theta^*}(x,0)=\infty$, and $\displaystyle\lim_{x\rightarrow 0} J_{\theta^*}(x,0)<0$ when $\frac{R_e}{\theta^*}e^{-\frac{\theta^*}{R_e}}<\frac{\tau}{1-\tau}$.
\item For $c_r<\frac{1}{1-\tau}$, $q(x)$ is monotonically decreasing in $x$ with $\displaystyle\lim_{x\rightarrow 0} q(x)>0$ and $\displaystyle\lim_{x\rightarrow \infty} q(x)=-\infty$.
\end{enumerate}
\end{lemma}
\IEEEproof See Appendix~\ref{app:h_fn}.

{\bf Remarks:} Observe that the above conditions are stated in terms of the design variables (e.g., $\tau$ and $c_r$).
It is clear that $R_e\le\theta^*,$ since $R_e$ is the relative gain due to rate refinement and cannot be greater than the optimal throughput $\theta^*.$ It follows that
 $\frac{R_e}{\theta^*}e^{-\frac{\theta^*}{R_e}} <e^{-1}.$ For example, to satisfy the condition in i), it would suffice to have $\tau>1/(1+\exp(1))$.
This range of $\tau$ is applicable to many scenarios (e.g., IEEE 802.11).
Moreover, it is easy to satisfy the condition in ii) by choosing $c_r\le 1/(1-\tau)-\delta,$ where $\delta>0$.
\begin{figure*}[ht]
\begin{tabular}{cc}
\begin{minipage}{.5\textwidth}
\includegraphics[width=0.8\textwidth]{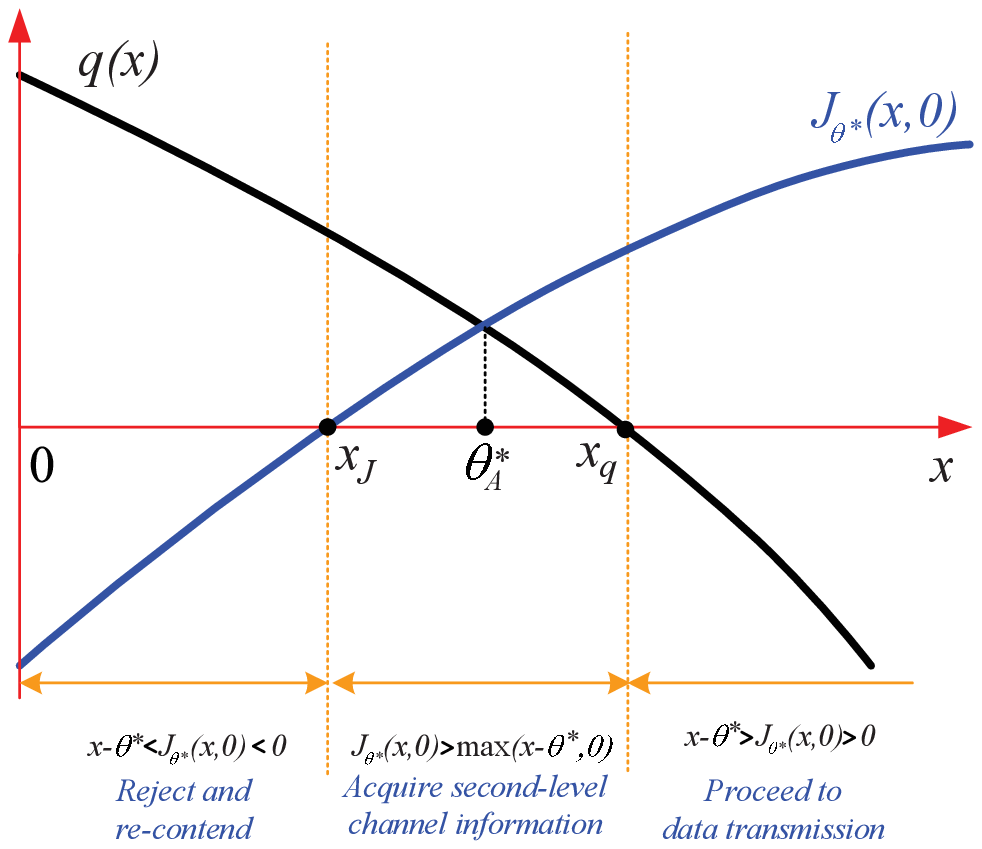}
\caption{A structural sketch for Strategy A.} \label{fig:fig1}
                \end{minipage}
&
\begin{minipage}{.5\textwidth}
\includegraphics[width=0.9\textwidth]{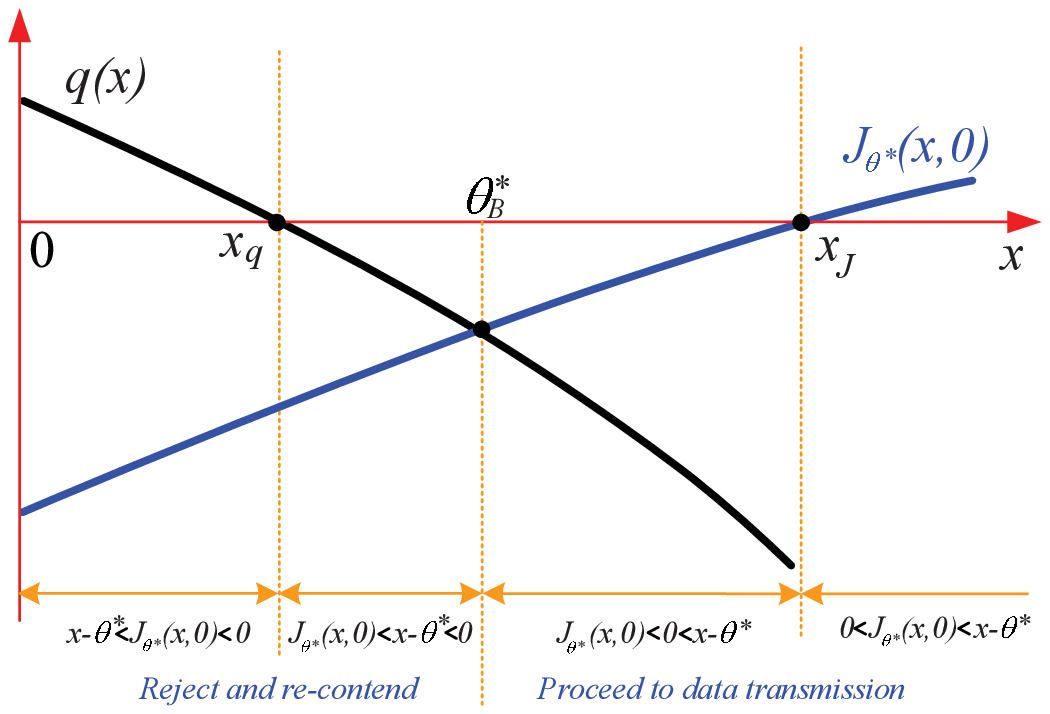}
\caption{A structural sketch for Strategy B.} \label{fig:fig2}
    \end{minipage}
    \end{tabular}
\end{figure*}

\renewcommand{\theenumi}{\roman{enumi}}
\begin{lemma}\label{opt_soln}
There exists at most one solution, in terms of {\small$\{x_J,x_q,\theta^*\},$} to the following system of equations:
\begin{equation}
\label{eq:optimal_strategy_1}
\left\{
\begin{array}{l}
\int_{{\theta^*}}^\infty (1-G(u|x_J))du =\frac{\theta^*\tau}{1-\tau},\\
(c_r(1-\tau)-1)x_q+(1-\tau)\left(R_e+\int_{0}^{{\theta^*}} G(u|x_q)du\right) =0,\\
\int_{x_J}^{x_q}J_{\theta^*}(u,0)\,dF(u)+\int_{x_q}^\infty (u-\theta^*)\,dF(u)=\frac{\theta^*\tau}{p_s}.
\end{array}
\hspace{-3mm}\right.
\end{equation}
\end{lemma}
%
%
Recall that $x_J$ and $x_q$ are the solutions to $J_{\theta^*}(x,0)=0$ and $q(x)=0$, respectively. From Lemma~\ref{h_fn}, it is easy to see that there is at most one pair $\left\{x_J,x_q\right\}$ satisfying (\ref{eq:optimal_strategy_1}). Similarly, since $J_{\theta^*}(x,0)$ and $q(x)$ intercept at $x={\theta^*}$, there exists at most one $\theta^*$ due to the monotonic nature of $J_{\theta^*}(x,0)$ and $q(x)$.

For convenience, let $\{x_J, x_q, \theta_A^*\}$ denote the solution to~(\ref{eq:optimal_strategy_1}) with $x_J \le x_q$, and ${\theta_B^*}$ be the solution to~(\ref{optimality_2}). Using the above lemmas, we obtain the following result on the structure of optimal scheduling strategy.
\begin{theorem}
\label{struct}
{\em
The optimal strategy for DOS with two-level probing, takes one of the two forms:
\newline
{\bf[Strategy A]}  It is optimal for the successful link
\renewcommand{\theenumi}{\roman{enumi}}
\begin{enumerate}
\item{ to transmit immediately after the first-level probing if $R_n^{(1)} > x_q$; } or
\item{ to give up the transmission and let all links re-contend if $ R_n^{(1)}< x_J$;} or
\item{ to engage in second-level probing if $R_n^{(1)}\in\left[x_J,x_q\right]$;
upon computing the new rate $R_n^{(2)}$, transmit at rate $R_n^{(2)}$ if $R_n^{(2)}>\theta_A^*$, or to give up the transmission otherwise.}
\end{enumerate}
Furthermore, the throughput under Strategy A is $\theta_A^*$. \newline
{\bf[Strategy B]} There is never a need to perform second-level probing. That is, it is optimal for the successful link to transmit at the current rate $R_n^{(1)}$ if $R_n^{(1)}>\theta_B^*$, or to defer its transmission and re-contend otherwise.
Furthermore, the throughput under Strategy B is $\theta_B^*$.
}
\end{theorem}
\proof See Appendix~\ref{app:struct}.

\subsection{Optimality Conditions}
In previous sections, we have studied DOS with two-level probing within the OST framework,
and  characterized the structure of optimal scheduling strategies. Our findings reveal that optimal scheduling may take either of two forms:  Strategy A or Strategy B. The next key step is to determine the conditions under which it is optimal to use Strategy A or Strategy B. We show that this can be easily determined by performing a threshold test on the function $J_{\theta^*}(\cdot,\cdot)$. We have the following theorem.

\begin{theorem}
\label{optimality}
\renewcommand{\theenumi}{\Alph{enumi}}
{\em
Strategy A is optimal if $J_{\theta_A^*}\left(\theta_A^*,0\right)\ge0$;
else, Strategy B is optimal.
}

\end{theorem}
\proof See Appendix~\ref{app:optimality}.

\subsection{Numerical Results}
\label{sec:numerical}
\begin{figure}[htp]
\includegraphics[width=0.5\textwidth]{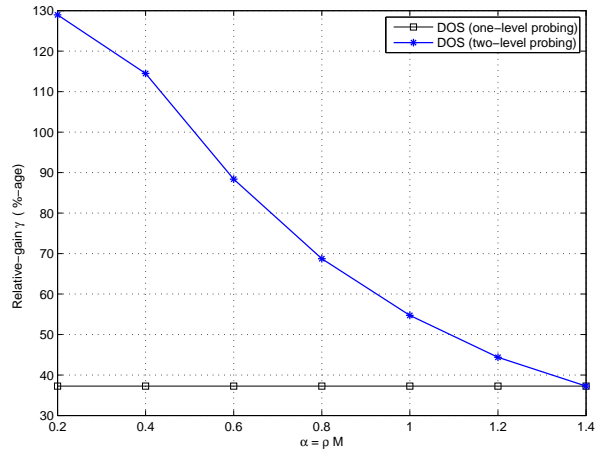}
\caption{ Relative gain $\Gamma$ as a function of $\alpha=\rho M$.} \label{fig:throughput_gain_plot}
\end{figure}

In this section, we provide a numerical example to illustrate the effectiveness of the proposed DOS with two-level probing under noisy estimation. Specifically, we compare the performance of the proposed {\em DOS with two-level probing}, with  that of {\em DOS with one-level probing} and   {\em PHY-oblivious scheduling}. The baseline for comparison is the PHY-oblivious scheduling that does not make use of any link-state information.
 We focus on the {\em relative gain} over PHY-oblivious scheduling, which is a function of $\rho M$, and is defined as
\[\Gamma(\rho M)= \frac{\theta-\theta_L}{\theta_L}.\]
 We set $p_s=\exp(-1)$, $M=300$ and $W=3000$, so that $\tau_t=0.1$ and $\tau=0.2$.
 Fig.~\ref{fig:throughput_gain_plot} depicts the performance comparison.
 It is clear that the relative gain achieved by DOS with two-level probing substantially outperforms that obtained by DOS with one-level probing. Observe that  the performance gain is significant in the low SNR regime (i.e., smaller values of $\alpha$). As $\alpha$ increases, the relative gain of DOS with two-level probing approaches that of DOS with one-level probing, and our intuition is that, for higher values of $\alpha$, the cost of overhead offsets that of the rate gain due to additional probing. Accordingly, the ``gray area'' between two thresholds ($x_h$ and $x_q$) collapses, and the optimal strategy degenerates to Strategy B, which is essentially DOS with first-level probing.

\section{DOS with Two-Level Probing: The Limited Feedback Case}
\label{limited_feedback}
In the above studies, it is assumed that for the link with successful contention,
its transmitter has the knowledge of the rate estimate for data transmissions.
In some practical scenarios, there is only limited feedback from the receiver to the transmitter.
With this motivation, we extend the study on DOS with two-level probing,
to the case where the feedback from the receiver to its transmitter takes the form $(0,1,e)$. More specifically, the decisions  from the receiver to the transmitter are conveyed by using ``NACK/ACK/ERASURE" signaling, where ``NACK"  is represented by ``0'' corresponding to the decision of {\em defer and re-contend}, ``ACK'' by ``1''  corresponding to the decision of {\em transmit}, and ``ERASURE'' by ``e''  indicating that the rate estimate falls in the gray area.

\subsection{One-level probing}
We first consider DOS with one-level probing, with one-bit feedback from  the receiver to its transmitter.
The basic idea is as follows. A constant transmission rate, denoted as
$R_1$, is pre-determined and known to the transmitter, and the data transmission takes place only when the one-bit feedback is ``1''.
A central problem here is to design the transmission strategy for maximal throughput.
Let $\gamma$ be the price function per unit time. Then, given its current rate estimate  $R_n^{(1)}$, the successful link in the the $n-$th probing has two options:
\begin{itemize}
\item ``1''--- transmit at rate $R_1$, and the corresponding reward is $R_1I(R_n^{(1)}>R_1)-\gamma$;
\item ``0''--- defer and re-contend, with the expected reward of $r_0$.
\end{itemize}
Clearly, there is an average cost of $\gamma{\tau}/{p_s}$ for every successful contention.

Let {\small$\displaystyle{{\hat \gamma}\triangleq \sup_{N\in Q}{E[R_NT]}/{E[T_N]}}$} be the optimal throughput. Then,
based on Theorem~\ref{exist},   the optimality equation is given by
\begin{equation}
\label{optimal_1_1bit}
E\left[R_1I(R^{(1)}>R_1)-{\hat \gamma}\right]^+=\frac{{\hat \gamma}\tau}{p_s}.
\end{equation}
As a result, we can show that the optimal policy in this case still has a threshold structure
with $R_1$ being the threshold. Furthermore, noting that $R_n^{(1)}\sim \exp({E[R^{(1)}]})$,
we conclude that the average throughput is given as
\[{\hat \gamma}= \frac{R_1e^{-\frac{R_1}{{E[R^{(1)}]}}}}{\frac{\tau}{p_s}+e^{-\frac{R_1}{{E[R^{(1)}]}}}}.\]

Observe that ${\hat \gamma}$ is a function of $R_1$. For a given stopping rule,
  $R_1$ can be chosen to  maximize the throughput, i.e., the optimal transmission rate ${\hat R_1}$ and the corresponding throughput obey that
\[{\hat R}_1=\argmax_{R_1}{\hat\gamma};\mbox{ and } {\hat \gamma}_{max}={\hat \gamma}({\hat R}_1).\]
It can be shown that ${\hat R_1}$ is  the solution to
\begin{equation}
\label{max_r_1}
\left(\frac{R_1}{E[R^{(1)}]}-1\right)e^{\frac{R_1}{E[R^{(1)}]}}=\frac{p_s}{\tau}.
\end{equation}

It follows that the optimal throughput is given by
\begin{equation}
\label{max_1_bit}
{\hat \gamma}_{max}={\hat R}_1-E[R^{(1)}]=\frac{p_s}{\tau}E[R^{(1)}]e^{\frac{-{\hat R}_1}{E[R^{(1)}]}}.
\end{equation}
\subsection{Two-level probing}
Next, we study DOS with two-level probing, with the feedback taking the form of $(0,1,e)$.
Along the same line as in the studies in Section~\ref{sec:chanest}, the receiver of the successful link, depending on its rate estimate $R_n^{(1)}$, presents three options to its transmitter:
\begin{itemize}
\item ``$1$''--- transmit at the rate $R_1$;
\item ``$0$''--- defer and re-contend;
\item ``$e$''--- perform a second-level probing to obtain $R_n^{(2)}$, and then decide:
\begin{itemize}
\item ``$1$''--- to transmit at rate $R_1$;
\item ``$0$''--- defer and re-contend.
\end{itemize}
 \end{itemize}
Define {\small$\displaystyle{\gamma^*=\sup_{N\in Q}{E[R_NT_{d,N}]}/{E[T_N]}}$}, which represents the optimal throughput for the given $R_1$. By Theorem~\ref{exist}, this corresponds to $r_0=0$. Since, $\gamma^*$ is the function of the rate $R_1$, we further maximize the throughput over all choices of $R_1$, by defining $\displaystyle{\gamma^*_{max}=\max_{R_1} \gamma^*}$.

We can write the expected net reward function corresponding to the second-level probing as
\[V_{\gamma^*}(x,R_1)=(1-\tau)(R_1-\gamma^*)\int_{R_1}^\infty G(dy|x)-\gamma^*\tau,\]
which can be further simplified as
\begin{eqnarray*}
V_{\gamma^*}(x,R_1)&=&(1-\tau)(R_1-\gamma^*)\left(1-G\left(R_1|x\right)\right)-\gamma^* \tau.
\end{eqnarray*}
The optimality equation in this case is given by
{\small
\begin{equation}\label{eq:optimality_1bit_2level}
E\left[\max\left\{R_1I\left(R^{(1)}\ge R_1\right)-\gamma^*,V_{\gamma^*}(R^{(1)},R_1)\right\}\right]^+=\gamma^*\frac{\tau}{p_s}.
\end{equation}
}
The following lemma gives useful bounds on the optimal throughput.
\begin{lemma}
{\em For a given transmission rate $R_1$, the optimal throughput obeys that
\[\gamma_L\le \gamma^* \le \gamma_U, \]
where
 \[\gamma_L\triangleq \frac{(1-\tau)R_1}{(1-\tau)+\tau\left(1+\frac{1}{p_s}\right)e^{\frac{R_1}{E[R^{(2)}]}}};\;\;
 \gamma_U\triangleq \frac{R_1}{1+\frac{\tau}{p_s}}.\;\;
  \]
  }
\end{lemma}
\noindent
{\bf Remarks:\\}
a) The lower bound $\gamma_L$ is obtained by using a strategy
 where the successful link always performs a second level probing, and  then decides to transmit for a duration
 of $1-\tau$ or to re-contend. \\
b) The upper bound $\gamma_U$ is achieved by a genie aided scheme, where the successful link contends only when its channel is good and there is no transmission outage.

Next we turn our attention to  structural properties of the optimal strategy. For convenience, define the relative gain function as
\[q_{\gamma^*}(x,R_1)\triangleq V_{\gamma^*}(x,R_1)-R_1I(x\ge R_1)+\gamma^*.\]

\begin{lemma}\label{v_fn}
{\em
$V_{\gamma^*}(x,R_1)$ and $q_{\gamma^*}(x,R_1)$ are characterized by the following properties:
\begin{enumerate}
\item $V_{\gamma^*}(x,R_1)$ is monotonically increasing in $x$. Furthermore, $\displaystyle\lim_{x\rightarrow\infty} V_{\gamma^*}(x,R_1)=c_1>0$, if $\tau \le 1-p_s$, and $\displaystyle\lim_{x\rightarrow 0} V_{\gamma^*}(x,R_1)<0$, when
$\tau \ge 0.5(\ln(1+\frac{1}{p_s})-1)$.
\item $q_{\gamma^*}(x,R_1)\ge 0$ for $x<R_1$; and $q_{\gamma^*}(x,R_1)<0$ for $x\ge R_1$.
\end{enumerate}
}
\end{lemma}
\IEEEproof See Appendix~\ref{app:v_fn}.

\begin{figure*}[tb]
\newcounter{mytempeqncnt}
\setcounter{mytempeqncnt}{\value{equation}}
\setcounter{equation}{29}
\normalsize
\begin{eqnarray}
\label{eq:optimal_strategy__1bit_1}
&&\gamma^*=\frac{(1-\tau)R_1\int_{x_v}^{R_1}(1-G(R_1|u))dF(u)+R_1e^{-\frac{R_1}{E[R^{(1)}]}}}{(1-\tau)\left(\int_{x_v}^{R_1}(1-G(R_1|u))dF(u)+e^{-\frac{R_1}{E[R^{(1)}]}}\right)+\tau(\frac{1}{p_s}+ e^{-\frac{x_v}{E[R^{(1)}]}})}\\\label{eq:optimal_strategy__1bit_2}
&&(1-\tau)(R_1-\gamma^*)\left(1-G\left(R_1|x\right)\right)=\gamma^* \tau.
\end{eqnarray}
\setcounter{equation}{31}
\hrulefill
\end{figure*}

The above  lemma serves as the basis to determine the optimal DOS scheduling under the feedback of $(0,1,e)$.
 Specifically, from the properties of $V_{\gamma^*}(\cdot,R_1)$, there exists some $x_v$ such that
 \begin{equation} \label{eq:v_fn}
 V_{\gamma^*}(x,R_1) \ge 0,\;\forall x\ge x_v,
 \end{equation}
 which, in turn, gives a threshold below which the option of ``defer and re-contend'' is optimal. From the properties of $q_{\gamma^*}(\cdot,R_1),$ it is also clear that for all $x\ge R_1$, it is optimal to transmit immediately without a second-level probing. Therefore, the interval $[x_v,R_1]$ defines the gray area where one could benefit from performing a second-level probing.

  We note that the throughput, denoted by $\gamma^*$, is the parameter to be optimized over the thresholds $x_v$ and $R_1$.
Combining~(\ref{eq:optimality_1bit_2level})~and~(\ref{eq:v_fn}), we establish ~(\ref{eq:optimal_strategy__1bit_1})~and~(\ref{eq:optimal_strategy__1bit_2}) that relate three key parameters, namely the lower threshold $x_v$, the transmission rate $R_1$, and the throughput ${\gamma^*}$.
It can be seen from~(\ref{eq:optimal_strategy__1bit_1})~and~(\ref{eq:optimal_strategy__1bit_2}) that $x_v=f_1(R_1,\gamma^*)$ and $\gamma^*=f_2(x_v,R_1)$, indicating that $\gamma^*=g(R_1)$. Then, $R_1$ can be chosen to the one maximizing $g(R_1)$,
i.e.,
\[ R_1^*=\argmax_{R_1} g(R_1);\mbox{ and } \gamma^*_{max}=g(R_1^*).
\]
Accordingly, the optimal $x_v^*$ is given by
\[
x_v^*= f_1(R_1^*,\gamma_{max}^*) .
\]

 Let $\{x_v^*,R_1^*,\gamma^*_{max}\}$ be the set of parameters obtained as outlined above. Also, let ${\hat R}_1$ be the solution to~(\ref{max_r_1}). The optimal strategy in the case with limited feedback is given by the following result.
\begin{theorem}
\label{strategy_csir}
{\em
The optimal strategy for DOS with two-level probing, with $(0,1,e)$ feedback, takes one of the two forms:
\newline
{\bf[Strategy A]}  It is optimal for the receiver of the successful link to
\renewcommand{\theenumi}{\roman{enumi}}
\begin{enumerate}
\item{feed back ``1'' if $R_n^{(1)} \geq R_1^*$, indicating to transmit at rate $R_1^*$ immediately after the first-level probing; or}
\item{feed back ``0'' if $ R_n^{(1)}< x_v^*$, indicating  to give up the transmission and let all links re-contend;}  or
\item feed back ``e'' if $R_n^{(1)}\in\left[x_v^*,R_1^*\right)$, indicating to engage in second-level probing; and upon computing the new rate $R_n^{(2)}$,
\begin{enumerate}
\item feed back ``1'' if $R_n^{(2)}\ge R_1^*$, indicating to transmit at rate $R_1^*$;  or
\item  feed back ``0'' if $R_n^{(2)}<R_1^*$, indicating to give up the transmission and re-contend.
 \end{enumerate}

\end{enumerate}
Furthermore, the throughput under Strategy A is $\gamma^*_{max}$. \newline
{\bf[Strategy B]} There is never a need to perform second-level probing. That is, it is optimal for receiver of the successful link to
\begin{enumerate}
\item feed back ``1'' if $R_n^{(1)}\ge{\hat R}_1$, indicating to transmit at rate ${\hat R_1}$;  or
\item  feed back ``0'' if $R_n^{(1)}< {\hat R}_1$, indicating to give up the transmission and re-contend.
 \end{enumerate}
Furthermore, the throughput under Strategy B is ${\hat \gamma}$.
}
\end{theorem}
\IEEEproof
The proof follows the same line of that for Theorem~\ref{struct}.


\section{Conclusion}\label{sec:conclusion}
We have considered distributed opportunistic scheduling for single-hop ad-hoc networks in which many links contend for the same channel using random access. Specifically, we have investigated DOS with two-level channel probing by optimizing the tradeoff between the throughput gain from more accurate rate estimation and the corresponding probing overhead. Capitalizing on optimal stopping theory with two-level incomplete information, we have showed that the optimal scheduling policy is threshold-based and is characterized by either one or two thresholds, depending on system settings. We have also identified optimality conditions. In particular, our analysis reveals that DOS with second-level channel probing is optimal when the first-level estimated rate falls in between the two thresholds.
By a numerical example, we have illustrated the effectiveness of the proposed DOS with two-level channel probing.
Finally, we considered the extension of DOS with two-level probing to the case where there is a limited feedback, of the form $(0,1,e)$, from the receiver to its transmitter.

So far, we have considered DOS with two-level probing, where we assumed that the refinement of the rate estimate is carried out once, via second-level probing of duration $\tau$.
 However, we can further extend this to $L-$level probing, where for $k=1,\ldots,L-1$ has the options 1) to transmit, or 2) to defer and re-contend, or 3) to resort to $(k+1)-$st level training at the cost of additional overhead. It is of great importance to devise well-structured, yet simple policies.

 We note that the proposed distributed scheduling with two-level probing  provides a new framework to study joint PHY/MAC optimization in practical networks where noisy probing is often the case and imperfect information is inevitable. We believe that this study provides some initial steps towards opening a new avenue on exploring channel-aware distributed scheduling for ad-hoc networks to enhance spectrum utilization; and this is potentially useful for enhancing MAC protocols for wireless local area networks (LANs) and wireless mesh networks.



\appendices
\renewcommand{\theenumi}{\arabic{enumi}}
\section{Derivation of Rate Equation~(\ref{eq:RM})}\label{rate}

Let $\beta^{(1)} \triangleq E\left[|\tilde{h}^{(1)}|^2\right]$, we follow the approach proposed in \cite{vakili06} and normalize $|\hat{h}^{(1)} |^2$ and $|\tilde{h}^{(1)} |^2$ as
\begin{eqnarray}
\hat{\lambda}^{(1)} &=& \frac{|\hat{h}^{(1)}|^2}{1-\beta^{(1)} }, \\
\zeta^{(1)} &=& \frac{|\tilde{h}^{(1)}|^2}{\beta^{(1)} },
\end{eqnarray}
where both $\hat{\lambda}^{(1)}$ and $\zeta^{(1)}$ are exponential-distributed with unit variance.

Defining the ``effective channel SNR'' and ``normalized error variance'' as
\begin{eqnarray}
\rho^{(1)}_{eff}&\triangleq& (1-\beta^{(1)})\rho,\label{eq:rhoM}\\
\alpha^{(1)} &\triangleq& \frac{\beta^{(1)}}{1-\beta^{(1)}}\label{eq:alphaM},
\end{eqnarray}
respectively. Substituting (\ref{eq:rhoM}) and (\ref{eq:alphaM}) in (\ref{eq:acsnr}) results in
\begin{equation} \label{eq:nacsnr}
\lambda^{(1)} = \frac{\rho^{(1)}_{eff}\hat{\lambda}^{(1)}}{1+\alpha^{(1)}\rho^{(1)}_{eff}\zeta^{(1)}}.
\end{equation}

It has been shown in \cite{vakili06} that the conditional probability distribution function (pdf) of $\lambda^{(1)}$ given $\hat{\lambda}^{(1)}$ takes the following form

{\small
\begin{eqnarray}\nonumber
f\left(\left.\lambda^{(1)} \right| \hat{\lambda}^{(1)}\right) &=&
\frac{\hat{\lambda}^{(1)}}{\alpha^{(1)} \left[\lambda^{(1)}\right]^2}
\exp\left\{-\frac{1}{\alpha^{(1)}}\left(\frac{\hat{\lambda}^{(1)}}{\lambda^{(1)}}
-\frac{1}{\rho^{(1)}_{eff}}\right)\right\}\\  \label{eq:conlam}
&&\mathbf{I}\left(\frac{\hat{\lambda}^{(1)}}{\lambda^{(1)}}\ge\frac{1}{\rho^{(1)}_{eff}}\right),
\end{eqnarray}
}
where $\mathbf{I}(\cdot)$ is the indicator function.

The following linear backoff function is employed to prevent channel outage.
\begin{equation} \label{eq:backoff}
\lambda_c(\hat{\lambda}^{(1)}) = \sigma_M \rho_{eff}\hat{\lambda}^{(1)},
\end{equation}
where $\sigma_M$ is the backoff factor with $0<\sigma_M<1$.

Define the optimal stopping rule over the $\sigma$-field $\mathcal{F}'$ generated by
$\{(\rho|\hat{h}_j|^2, K_j), j=1,2,\ldots,n\}$. Using (\ref{eq:conlam}) and (\ref{eq:backoff}), the conditional expectation ${R}_n^{(1)}$ can be computed as

{\small
\begin{eqnarray*}
{R}_n^{(1)} & =& E\left[\log\left(1+\lambda_c(\hat{\lambda}^{(1)})\right)\mathbf{I}\left(\lambda_c(\hat{\lambda}^{(1)})
\le \lambda^{(1)}\right)|\mathcal{F}'\right], \\
&=&\left[1-\exp\left\{\frac{-\left(\frac{1}{\sigma_M}
- 1\right)}{\alpha^{(1)}\rho^{(1)}_{eff}}\right\}\right] W\log\left(1+\sigma_M
\rho|\hat{h}^{(1)} |^2\right).
\end{eqnarray*}
}
For the low SNR wideband regime where $\rho\rightarrow \infty $ and $W=\Theta(\frac{1}{\rho})$,  $\mbox{ }{R}_n^{(1)} $ can be well approximated by
\begin{equation}\label{eq:approxR}
{R}_n^{(1)}  \approx \rho W\sigma_M |\hat{h}^{(1)} |^2.
\end{equation}
\section{Proof of Lemma~\ref{exist} }\label{app:exist}
{For a given $\theta$, let $N(\theta)$ be a stopping rule such that
{\small
\begin{eqnarray*}
N(\theta)=\arg\sup_{N\in {Q}}E\left[ R_NT_{d,N}-\theta T_N \right].
\end{eqnarray*}
}
Let {\small $Z_n\stackrel{\Delta}{=}R_nT_{d,n}-\theta T_n$}. Then, it follows from Theorem~1 in~\cite[Chapter 3]{ferguson:os} that $ N(\theta)$ exists if the following conditions are satisfied:
\[\mbox{(A1) }E[\sup_n Z_n]< \infty, \mbox{ and } \mbox{(A2) }\limsup_{n\rightarrow\infty}Z_n=-\infty, \mbox{ a.s.,}
\]
Since, it is clear that $\displaystyle{\limsup_{n\rightarrow \infty}Z_n=-\infty}$, we can easily verify (A2).

For some $0<\mu<{1}/{p_s}$, we introduce
{
\[Z_n'= \max\{R_n^{(1)},R_n^{(2)}\}T-n(\theta \frac{\tau}{p_s}-\mu)\]
{ and }
\[Z_n''=\sum_{j=1}^n \left(\frac{1} {p_s}-K_j-\mu\right).\]
Then, we note that
\[E\left[\sup_nZ_n\right]\le E\left[\sup_n Z_n'\right]+E\left[\sup_n Z_n''\right]\]
Appealing to Theorem~1 and Theorem~2 of~\cite[Chapter 4]{ferguson:os}, we conclude that
{\small $E\left[Z_n'\right]<\infty \mbox{ and }E\left[Z_n''\right]<\infty$}, respectively.
Therefore (A1) holds.

The second part of the lemma follows directly from Theorem 1 in ~\cite[Ch.6]{ferguson:os}.
}

\section{Proof of Lemma~\ref{h_fn}}\label{app:h_fn}
\noindent a) Using Fubini's theorem, we rewrite $J_{\theta}(x,r)$ as
\begin{equation}
J_{\theta^*}(x,0) = (1-\tau) \int_{{\theta^*}}^\infty
(1-G(u|x))du - \theta^*\tau.
\end{equation}

Since $G(y|x)$ decreases monotonically with $x$, $J_{\theta^*}(x,0)$ is also monotonically increasing in $x$.
Note that $\displaystyle\lim_{x\rightarrow \infty}(1-G(u|x))=1$. Then, by Lebesgue's convergence theorem, we have  $\lim_{x\rightarrow \infty}J_{\theta^*}(x,0)=\infty$. Next, recall that $|z|^2 \sim \exp(R_e)$. It follows that
{\small $\displaystyle\lim_{x\rightarrow 0}G(y|x)=G_{|z|^2}(y)=1-e^{-\frac{y}{R_e}}$,}
and consequently,
\begin{eqnarray}
\lim_{x\rightarrow 0}J_{\theta^*}(x,0)
&=&(1-\tau)R_e e^{-\frac{{\theta^*}}{R_e}}-\theta^*\tau.
\end{eqnarray}

Thus, under the condition {\small $R_e/\theta^* e^{-\frac{{\theta^*}}{R_e}}<\tau/(1-\tau)$},
\begin{equation}
\lim_{x\rightarrow 0}J_{\theta^*}(x,0)<0.\label{eq:h-lessthan0}
\end{equation}

\noindent b) Using Fubini's theorem, we can rewrite $J_{\theta^*}(x,r)$ as
{\small
\begin{equation}
\label{fub_q}
J_{\theta^*}(x,0) = (1-\tau) \left(c_r x+R_e-\theta^*+\int_{0}^{{\theta^*}}\hspace{-3mm} G(u|x)du \right)-\theta^*\tau
\end{equation}
}
It follows that
{\small
\[q(x)=(c_r (1-\tau)-1)x+(1-\tau)R_e+(1-\tau)\int_0^{{\theta^*}} \hspace{-3mm}G(u|x)du.\]
}
We can verify that
{\small
\[\lim_{x\rightarrow 0}q(x)=R_e(1-\tau)+(1-\tau)\theta^*>0\]
}
Furthermore, when {\small$c_r< \frac{1}{1-\tau}$,} it is clear that
{\small \[\lim_{x\rightarrow\infty}q(x)=-\infty.\]}
Since $G(y|x)$ is monotonically decreasing in $x$, we conclude that $q(x)$ is also monotonically decreasing in $x$.

\section{Proof of Theorem~\ref{struct}}\label{app:struct}
$\mbox{ Let }x_J\mbox{ and }x_q \mbox{  be solutions to }J_{\theta^*}(x,0)=0$ and $q(x)=0$ respectively. From Lemma~\ref{h_fn}, we have
{\small
\begin{equation}\label{eq:xh}
J_{\theta^*}(x,0)\left\{
\begin{array}{ll}
<0&\mbox{ if } x<x_J\\
=0&\mbox{ if } x=x_J\\
>0&\mbox{ if } x>x_J
\end{array}
\right.
\end{equation}
}
and
{\small
\begin{equation}\label{eq:xq}
q(x)\left\{
\begin{array}{ll}
<0&\mbox{ if } x>x_q\\
=0&\mbox{ if } x=x_q\\
>0&\mbox{ if } x<x_q.
\end{array}
\right.
\end{equation}
}
Thus, one of the following two possibilities holds.
\begin{enumerate}
\item \underline {The case with $x_q \ge x_J$:}\\
 From the above discussions and the monotonicity properties of $J_{\theta^*}(\cdot,0)$ and $q(\cdot)$, it follows that
{\small \begin{equation}
\label{eq:max_Jx}
\max\left[x-\theta^*,J_{\theta^*}(x,0)\right]^+=\left\{
\begin{array}{ll}
x-\theta^*&\hspace{-4mm}\mbox{  if } x>x_q\\
J_{\theta^*}(x,0)&\hspace{-4mm}\mbox{  if } x\in[x_J,x_q]\\
0&\hspace{-4mm}\mbox{  if } x<x_J
\end{array}
\right.
\end{equation}
}
Furthermore, from~(\ref{eq:max_Jx}) and the optimality equation~(\ref{eq:optEmax2}), we have that
{\small \begin{equation}
\label{eq:opt_a}
 \int_{x_J}^{x_q}J_{\theta^*}(u,0)\,dF(u)+\int_{x_q}^\infty (u-\theta^*)\,dF(u)=\frac{\theta^*\tau}{p_s}.
\end{equation}
}
Consequently, it is clear that the optimal strategy is
{\small \begin{equation}\label{eq:phi}
\phi_n(R_n^{(1)})=\left\{
\begin{array}{ll}
1 \mbox{ (transmit)}&\mbox{ if } R_n^{(1)}>x_q\\
2\mbox{ (2-level)}&\mbox{ if } R_n^{(1)}\in[x_J,x_q]\\
0\mbox{ (re-contend)}&\mbox{ if } R_n^{(1)}<x_J
\end{array}
\right.
\end{equation}
}
and when $\phi_n(R_n^{(1)})=2$, the strategy is
{\small \begin{equation}
\psi_n(R_n^{(2)})=\left\{
\begin{array}{ll}
1 \mbox{ (transmit)}&\mbox{ if } R_n^{(2)}\ge\theta_A^*\\
0\mbox{ (re-contend)}&\mbox{ if } R_n^{(2)}<\theta_A^*
\end{array}
\right.
\end{equation}
}\noindent
where $\theta_A^*$ is the solution to~(\ref{eq:opt_a}).
It can be seen that thresholds $x_J$ and $x_q$ are found as the solutions to $J_{\theta^*}(x,0)=0$ and $q(x)=0$ respectively.
Thus, {\small $\{x_J,x_q,\theta_A^*\}$} is the solution to the system~(\ref{eq:optimal_strategy_1}).
An illustration of Strategy A is depicted in Fig. \ref{fig:fig1}.\\
\item \underline { The case with $x_q < x_J$: }\\
From~(\ref{eq:xh})~and~(\ref{eq:xq}), we have
{\small
\begin{equation}
\label{eq:max_Jx_b}
\max\left[x-\theta^*,J_{\theta^*}(x,0)\right]^+=\left\{
\begin{array}{ll}
x-\theta^*&\mbox{ if } x\ge\theta^*\\
0&\mbox{ if } x< \theta^*
\end{array}
\right.
\end{equation}
}\noindent
and {\small $J_{\theta^*}(x,0) <\max\left[x-\theta^*,0\right]$}. Therefore, it is never optimal to perform second-level probing. From~(\ref{eq:max_Jx_b}) and the optimality equation~(\ref{eq:optEmax2}) we obtain
{\small \[
\int_{\theta^*}^\infty\left(x-\theta^*\right)dF(x)=\frac{\theta^*\tau}{p_s},
\]
}\noindent
which is equivalent to ~(\ref{optimality_2}). Thus from~(\ref{eq:max_Jx_b}), the optimal strategy is

{\small \begin{equation}\label{eq:phi_b}
\phi(R_n^{(1)})=\left\{
\begin{array}{ll}
1 \mbox{ (transmit)}&\mbox{ if } R_n^{(1)}\ge {\theta_B^*}\\
0\mbox{ (re-contend)}&\mbox{ if } R_n^{(1)}<\theta_B^*,
\end{array}
\right.
\end{equation}
}\noindent
where the threshold $\theta_B^*$ is the solution to~(\ref{optimality_2}).
An illustration of Strategy B is depicted in Fig. \ref{fig:fig2}.
\end{enumerate}

\section{Proof of Theorem~\ref{optimality}}\label{app:optimality}
Suppose {\small $J_{\theta_A^*}\left(\theta_A^*,0\right)\ge 0$}. Then, this implies that {\small $J_{\theta_A^*}\left(\theta_A^*,0\right)\ge \max[x-{\theta_A^*},0]$} when $x=\theta_A^*$.
Specifically, when {\small$R_1^{(1)}={\theta_A^*}$}, performing second-level probing and using an optimal strategy thereafter yield an expected reward of $J_{\theta_A^*}\left(\theta_A^*,0\right)$, which is at least as good as using Strategy B. Equivalently, we show that there exists at least one value of $x$ ($\theta_A^*$ in this case) for which performing second-level probing is optimal. We conclude that Strategy A is optimal.

Next, we assume Strategy A is optimal and show that $J_{\theta^*_A}(\theta_A^*,0)\ge 0$. Under such an assumption, there must exist some $x_1$ for which it is beneficial to demand additional information, i.e.
{\small \begin{equation}
\label{hmax}
 J_{\theta_A^*}(x_1,0)\ge\max[x_1-\theta_A^*,0].
\end{equation}
}
We now investigate $J_{\theta_A^*}(\theta_A^*,0)$ in two different cases, namely  $ \theta_A^*\ge x_1$ and $ \theta_A^*< x_1$.

\begin{enumerate}
\item \underline{ The case with $ \theta_A^*\ge x_1$:}\\
In this case,
{\small\begin{equation}
J_{\theta_A^*}(\theta_A^*,0)\ge J_{\theta_A^*}(x_1,0)\ge \max[x_1-\theta_A^*,0]=0,
\end{equation}
}\noindent
where the first and second inequalities are due to the monotonicity of $J(\cdot,0)$ and the assumed optimality of Strategy A, respectively.\\

\item \underline{ The case with $ \theta_A^*< x_1$:}\\
In this case,
\begin{equation}
J_{\theta_A^*}(\theta_A^*,0)\ge J_{\theta_A^*}(x_1,0)-x_1+\theta_A^*\ge 0,
\end{equation}
where the first inequality follows from the fact that $J_{\theta_A}(x,0)-x+\theta$ is decreasing in $x$ and the second inequality is due to (\ref{hmax}).
\end{enumerate}
Summarizing the above two cases, we conclude that $J_{\theta^*_A}(\theta_A^*,0)\ge 0$ is a necessary condition for the optimality of Strategy A. Using contra position, we conclude that Strategy B is optimal if {\small $J_{\theta^*_A}(\theta_A^*,0)< 0$}.

\section{Proof of Lemma~\ref{v_fn}}
\label{app:v_fn}
 It is clear that $V_{\gamma^*}(x,R_1)$ is monotonically increasing in $x$, and that
\[\displaystyle\lim_{x\rightarrow\infty}V_{\gamma^*}(x,R_1)= (1-\tau)R_1 -\gamma^* . \]
Since $\gamma^*\le \gamma_U$, it follows that $\displaystyle\lim_{x\rightarrow\infty}V_{\gamma^*}(x,R_1)>0$,
provided that $\tau \le 1-p_s$.
Furthermore, observe that
\begin{eqnarray*}
\lim_{x\rightarrow 0}V_{\gamma^*}(x,R_1)&=&(1-\tau)(R_1-\gamma^*)e^{-\frac{R_1}{R_e}}-\gamma^* \tau \\
&\le& \gamma^*\left((1-\tau)(\frac{R_1}{\gamma_L}-1)e^{-\frac{R_1}{R_e}}- \tau \right)\\
&=&  \tau \gamma^*\left((1+\frac{1}{p_s})e^{\frac{R_1}{E[R^{(2)}]}}e^{-\frac{R_1}{R_e}}-1\right)\\
&\le &\tau \gamma^*\left((1+\frac{1}{p_s})e^{-(1+2\tau)}-1\right),
\end{eqnarray*}
where the last inequality follows due to the fact that~$\frac{E[R^{(2)}]}{R_1}\le 1$ and $\frac{E[R^{(2)}]}{R_e}\approx (1+2\tau)$.
We conclude that
\[\lim_{x\rightarrow 0}V_{\gamma^*}(x,R_1)<0,\;\;\mbox{ for }\tau \ge 0.5\left(\ln\left(1+\frac{1}{p_s}\right)-1\right).\]

The second part follows from the facts that
for $x\ge R_1$,
{\footnotesize
\begin{eqnarray*}
\mbox{  }&q_{\gamma^*}(x,R_1)&\\
&=&\hspace{-0.6cm}(1-\tau)\left(1-G\left(R_1|x\right)\right)(\gamma^*-R_1)-\tau R_1< 0,
\end{eqnarray*}
}
and for $x < R_1$,
{\footnotesize
{\footnotesize
\begin{eqnarray*}
\mbox{  }&q_{\gamma^*}(x,R_1)&\\
&=&\hspace{-0.6cm}(1-\tau)R_1\left(1-G\left(R_1|x\right)\right)+(1-\tau)G\left(R_1|x\right)\gamma^*\ge 0.
\end{eqnarray*}
}

\bibliographystyle{IEEEtranS}
\bibliography{Bib}
\end{document}

%% file: pream.tex
\newtheorem{prop}{Proposition}[section]

\newtheorem{cor}{Corollary}

\newtheorem{lm}{Lemma}

\newtheorem{thm}{Theorem}

\newcommand{\bthm}{\begin{thm}}
\newcommand{\ethm}{\end{thm}}

\newcommand{\bcor}{\begin{cor}}
\newcommand{\ecor}{\end{cor}}
\newcommand{\bprop}{\begin{prop}}
\newcommand{\eprop}{\end{prop}}
\newcommand{\blm}{\begin{lm}}
\newcommand{\elm}{\end{lm}}
\newcommand{\beq}{\begin{equation}}
\newcommand{\eeq}{\end{equation}}
\newcommand{\ber}{\begin{eqnarray}}
\newcommand{\eer}{\end{eqnarray}}

\newenvironment{proof1}{\begin{trivlist}\item[]{\bf Proof:\hspace{2mm}}}{\hfill$\blackbox$\end{trivlist}}

\newcommand{\argmax}{\mathop{\mbox{\rm arg\,max}}}

\newcommand{\blackbox}{\vrule height7pt width5pt depth1pt}

\newcommand{\bit}{\begin{itemize}}
\newcommand{\eit}{\end{itemize}}
\newcommand{\ben}{\begin{enumerate}}
\newcommand{\een}{\end{enumerate}}
\newcommand{\bdesc}{\begin{description}}
\newcommand{\edesc}{\end{description}}
\newcommand{\beqarrn}{\begin{eqnarray*}}
\newcommand{\eeqarrn}{\end{eqnarray*}}
\newenvironment{proofof}[1]{\begin{trivlist}\item[]{\bf Proof of #1:\hspace{2mm}
}}{\hfill\blackbox\end{trivlist}}
\newcommand{\bproofof}{\begin{proofof}}
\newcommand{\eproofof}{\end{proofof}}
\newenvironment{rem}{\begin{trivlist}\item[]{\bf
Remark:}\hspace{4mm}}{\end{trivlist}}
\newcommand{\brem}{\begin{rem}}
\newcommand{\erem}{\end{rem}}
\newenvironment{rems}{\begin{trivlist}\item[]{\bf
Remarks}\begin{itemize}}{\end{itemize}\end{trivlist}}
\newcommand{\brems}{\begin{rems}}
\newcommand{\erems}{\end{rems}}
\newtheorem{fact}{Fact}
\newcommand{\bfact}{\begin{fact}}
\newcommand{\efact}{\end{fact}}
\newtheorem{examp}{Example}
\newcommand{\bexamp}{\begin{examp}\rm}
\newcommand{\eexamp}{\end{examp}}
\newtheorem{defn}{Definition}
\newcommand{\bdefn}{\begin{defn}\rm}
\newcommand{\edefn}{\end{defn}}

\newtheorem{prob}{Problem}
\newcommand{\bprob}{\begin{prob}}
\newcommand{\eprob}{\end{prob}}

\newcommand{\bvtm}{\begin{verbatim}}
\newcommand{\bfig}{\begin{figure}}
\newcommand{\efig}{\end{figure}}
\newcommand{\bcen}{\begin{center}}
\newcommand{\ecen}{\end{center}}

\long\def\comment#1{}

%% file: math_def.tex
\def \n2{{N_0 \over 2}}

\def \h5{\hspace{0.5in}}

%% file: ToNDraftJuly7.bbl
\begin{thebibliography}{10}
\providecommand{\url}[1]{#1}
\csname url@samestyle\endcsname
\providecommand{\newblock}{\relax}
\providecommand{\bibinfo}[2]{#2}
\providecommand{\BIBentrySTDinterwordspacing}{\spaceskip=0pt\relax}
\providecommand{\BIBentryALTinterwordstretchfactor}{4}
\providecommand{\BIBentryALTinterwordspacing}{\spaceskip=\fontdimen2\font plus
\BIBentryALTinterwordstretchfactor\fontdimen3\font minus
  \fontdimen4\font\relax}
\providecommand{\BIBforeignlanguage}[2]{{%
\expandafter\ifx\csname l@#1\endcsname\relax
\typeout{** WARNING: IEEEtranS.bst: No hyphenation pattern has been}%
\typeout{** loaded for the language `#1'. Using the pattern for}%
\typeout{** the default language instead.}%
\else
\language=\csname l@#1\endcsname
\fi
#2}}
\providecommand{\BIBdecl}{\relax}
\BIBdecl

\bibitem{belllab00}
M.~Andrews, K.~Kumaran, K.~Ramannan, A.~Stolyar, R.~Vijaykumar, and P.~Whiting,
  ``{CDMA} data {QoS} scheduling on the forward link with variable channel
  conditions,'' \emph{Bell Labs Tech. Memo.}, April 2000.

\bibitem{prophet_ineq}
D.~Assaf, L.~Goldstein, and E.~Samuel-Cahn, ``A statistical version of prophet
  inequalities,'' \emph{Annals of Statistics}, vol.~26, pp. 1190--1197, 1996.

\bibitem{bertsekas-gallager}
D.~P. Bertsekas and R.~Gallager, \emph{Data Networks}.\hskip 1em plus 0.5em
  minus 0.4em\relax Upper Saddle River, NJ: Prentice-Hall, 1992.

\bibitem{Borst:03}
S.~Borst, ``User-level performance of channel-aware scheduling algotirthms in
  wireless data networks,'' in \emph{Proc.\ IEEE \mbox{INFOCOM'03}}, San
  Francisco, CA, 2003.

\bibitem{ferguson:os}
T.~Ferguson, \emph{Optimal Stopping and Applications}.\hskip 1em plus 0.5em
  minus 0.4em\relax available at
  http://www.math.ucla.edu/\verb%~%tom/Stopping/Contents.html, 2006.

\bibitem{weiyan_interference}
W.~Ge, J.~Zhang, J.~Wieselthier, and S.~Shen, ``{PHY}-{Aware} distributed
  scheduling for ad hoc communications with physical interference model,''
  \emph{IEEE Transactions on Wireless Communications}, to appear, 2009.

\bibitem{ji-exploiting}
Z.~Ji, Y.~Yang, J.~Zhou, M.~Takai, and R.~Bagrodia, ``Exploiting medium access
  diversity in rate adaptive wireless {LANs},'' in \emph{Proc. ACM/IEEE
  MOBICOM'04}, Philadelphia, PA, Sept. 2004.

\bibitem{bl:kay}
S.~M. Kay, \emph{Fundamentals of statistical signal processing : estimation
  theory}.\hskip 1em plus 0.5em minus 0.4em\relax Englewood Cliffs, NJ:
  Prentice-Hall, 1993.

\bibitem{Xin:J02}
X.~Liu, E.~K. Chong, and N.~B. Shroff, ``A framework for opportunistic
  scheduling in wireless networks,'' \emph{Computer Networks}, vol.~41, no.~4,
  pp. 451--474, Mar. 2003.

\bibitem{qin03}
X.~Qin and R.~Berry, ``Exploiting multiuser dieversity for medium access
  control in wireless networks,'' in \emph{Proc. IEEE INFOCOM'03}, San
  Francisco, CA, Apr. 2003.

\bibitem{sad02}
B.~Sadeghi, V.~Kanodia, A.~Sabharwal, and E.~Knightly, ``Opportunistic media
  access for multirate ad hoc networks,'' in \emph{Proc. ACM/IEEE MOBICOM'02},
  Atlanta, GA, 2002.

\bibitem{Stadje97}
W.~Stadje, ``An optimal stopping problem with two levels of incomplete
  information,'' \emph{Mathematical Methods of Operations Research}, vol.~45,
  no.~1, pp. 119--131, 1997.

\bibitem{vakili06}
A.~Vakili, M.~Sharif, and B.~Hassibi, ``The effect of channel estimation error
  on the throughput of broadcast channels,'' in \emph{Proc. IEEE ICASSP'06},
  Toulouse, France, May 2006.

\bibitem{Tse02}
P.~Viswanath, D.~N.~C. Tse, and R.~Laroia, ``Opportunistic beamforming using
  dumb antennas,'' \emph{IEEE Trans. Info. Theory}, vol.~48, no.~6, pp.
  1277--1294, Jun. 2002.

\bibitem{bo_wideband}
B.~Wang, J.~Zhang, and L.~Zheng, ``Achievable rates and scaling laws of
  power-constrained wireless sensory relay networks,'' \emph{IEEE Trans. Inf.
  Theory}, vol.~52, pp. 4084--4104, 2006.

\bibitem{OSTAggregation}
Z.~Ye, A.~Abouzeid, and J.~Ai, ``Optimal stochastic policies for distributed
  data aggregation in wireless sensor networks,'' \emph{IEEE/ACM Trans. on
  Networking}, to appear, 2008.

\bibitem{zgz:mobihoc}
D.~Zheng, W.~Ge, and J.~Zhang, ``Distributed opportunistic scheduling for
  ad-hoc communications: An optimal stopping approach,'' in \emph{Proc. Mobihoc
  2007}, Montreal, Canada, Sept. 2007.

\bibitem{noisydos}
D.~Zheng, M.~O. Pun, W.~Ge, J.~Zhang, and H.~V. Poor, ``Distributed
  opportunistic scheduling for ad hoc communications with imperfect channel
  information,'' \emph{IEEE Transactions on Wireless Communications}, vol.~7,
  pp. 5450--5460, December 2008.

\end{thebibliography}
